\documentclass{aa}  

\usepackage{natbib}
\bibpunct{(}{)}{;}{a}{}{,} 

\usepackage{graphicx}
\usepackage{txfonts}
\usepackage{color}
\newcommand{\micron}{\mu{\rm m}}

\begin{document}

   \title{Seasonal variation of radial brightness contrast of Saturn's rings viewed in mid-infrared by Subaru/COMICS}


   \authorrunning{Fujiwara et al.}

   \author{Hideaki~Fujiwara\inst{1},
Ryuji~Morishima\inst{2,3},
Takuya~Fujiyoshi\inst{1}, 
\and 
Takuya~Yamashita\inst{4}, 
          }

   \institute{Subaru Telescope, National Astronomical Observatory of Japan, 
650 North A'ohoku Place, Hilo, HI 96720, USA\\
              \email{hideaki@naoj.org}
         \and
University of California, Los Angeles, Institute of Geophysics and Planetary Physics, Los Angeles, CA 90095, USA
         \and
Jet Propulsion Laboratory, California Institute of Technology, Pasadena, CA 91109, USA
         \and
National Astronomical Observatory of Japan, 2-21-1 Osawa, Mitaka, Tokyo 181-0015, Japan
             }

   \date{Received 9 October 2015 / Accepted xx December 2016}

 
  \abstract
   {}
   {To investigate the mid-infrared (MIR) characteristics of Saturn's rings. }
   {We collected and analyzed MIR high spatial resolution images of Saturn's rings obtained in January 2008 and April 2005 with COMICS mounted on Subaru Telescope, and investigated the spatial variation in the surface brightness of the rings in multiple bands in the MIR. We also composed the spectral energy distributions (SEDs) of the C, B, and A rings and the Cassini Division, and estimated the temperatures of the rings from the SEDs assuming the optical depths. }
   {We find that the C ring and the Cassini Division were warmer than the B and A rings in 2008, which could be accounted for by their lower albedos, lower optical depths, and smaller self-shadowing effect. We also find that the C ring and the Cassini Division were considerably brighter than the B and A rings in the MIR in 2008 and the radial contrast of the ring brightness is the inverse of that in 2005, which is interpreted as a result of a seasonal effect with changing elevations of the sun and observer above the ring plane. }
   {}


   \maketitle
%

\section{Introduction}
Ever since Saturn's rings were first observed by Galileo Galilei about 400 years ago, a multitude of scientists have been puzzling over their true characteristics and origin. Current understanding is that Saturn's rings consist of a set of geometrically thin disks containing cm- to m-sized particles covered mostly with water ice \citep[e.g.][]{french00}. In the 1980s, Voyager 1 and 2 revealed numerous remarkable fine structures in Saturn's rings including waves, spokes, and narrow rings. Voyager also discovered that the Cassini Division was not ``empty'' but there were some ringlets inside it. 

One of main theories regarding the origin of Saturn's rings is that the rings were formed by destruction of a giant moon \citep[e.g.][]{charnoz09,canup10}. This is very similar to the formation process of dusty debris disks around main-sequence stars. Dust grains in the disks are supplied through destruction of planetesimals or comets during planet formation process. Recent observations with high sensitivity infrared space telescopes including AKARI and Spitzer enable us to measure the characteristics of debris disks in the mid-infrared (MIR) wavelengths precisely \citep[e.g.][]{fujiwara12b, fujiwara13}. To utilize Saturn's rings as an experiment site for understanding of dust and particles in debris disks, MIR observations of Saturn's rings and analysis of their appearance with high spatial resolution is important. 

The first detection of the thermal emission of Saturn's rings in the MIR was reported by \citet{allen71}, followed by several ground-based observations using MIR photometers in the 1970s and 1980s \citep[e.g.][]{rieke75,nolt78,tokunaga80,froidevaux81a}. 
In the 2000s, a few ground-based imaging observations in the MIR were reported.
\citet{spilker03} and \citet{ferrari05} presented the images of Saturn's rings at 20.5~$\micron$ taken by CAMIRAS mounted on the 3.6-m Canada-France-Hawaii Telescope (CFHT) in July 1999 and March 2000. \cite{leyrat08b} added the 13.0 and 17.3~$\micron$ images taken by CFHT/CAMIRAS in July 1999 and the 19.5~$\micron$ image taken by VISIR mounted on the 8.2-m Very Large Telescope (VLT) in April 2005. Those MIR images were taken when the ring opening angle was large ($\gtrsim 20^\circ$), and the radial brightness contrasts of the Saturn's rings in the MIR images look similar to that in the visible light. 
In addition to ground-based observations, Spitzer also contributed to providing new infrared data of Saturn's rings. In 2009 Spitzer discovered an enormous, geometrically thick infrared ring around Saturn, extending from at least 128- to 207-times of Saturn's radius with an orbit tilted by $27^\circ$ from the main ring plane \citep{verbiscer09}. 

Recent observations made by Cassini using multiple onboard instruments have dramatically improved understanding of the structure of Saturn's rings \citep[e.g.][]{esposito11}. 
In particular, observations of the far-infrared (FIR) thermal emission from Saturn's rings by the Focal Plane 1 (FP1) channel, covering $17~\micron$--1~mm, of the Composite Infrared Spectrometer (CIRS) onboard Cassini allow us to investigate the physical properties of the ring material with high spatial resolution, since the wavelength range covers the peak of the blackbody emission of the ring material and therefore temperatures and filling factors could be measured simultaneously. 
\citet{spilker06} presented spatially-resolved FIR observations of thermal emission from the main rings using the FP1 channel of Cassini/CIRS in 2004 and 2005, and showed that particle temperatures decreased with increasing phase angle. 
\citet{morishima10} and \citet{morishima11} applied their new multilayer radiative transfer model \citep{morishima09} to the data in \citet{spilker06} and other azimuthally-scanned CIRS spectra including those in Saturn's shadow, and estimated radial profiles of albedo, the fraction of fast rotators, and the thermal inertia of the ring material. 
\citet{altobelli08} analyzed FP1 data of 65 radial scans of Saturn's rings and mapped temperatures and filling factors of ring particles in a multidimensional observation space. They extensively investigated the dependence of those values on solar phase angle, spacecraft elevation and solar elevation. 
\citet{leyrat08a} also investigated the dependencies of the temperature and the filling factor with the phase angle and the local hour angle using 48 azimuthal scans of rings by CIRS (FP1) and revealed a different thermal behavior between the A, B and C rings.
\cite{flandes10} analyzed FP1 data of Saturn's rings, derived the temperature of the main rings from a wide variety of geometries, and confirmed that the temperature of Saturn's main rings varied with the solar elevation angle. 
\citet{altobelli07} is the only published paper using the MIR data by the Focal Plane 3 (FP3, covering 9--17~$\micron$) channel of CIRS, which derives radial temperature profiles for the C ring with a higher spatial resolution and reveals fine structures in the ring. 

As is obvious from the results from the Cassini/CIRS observations, continuous observational efforts made at various epochs, geometries, and wavelengths are crucial in further understanding the structure and physical properties of Saturn's rings. 
Even in the Cassini era, ground-based telescopes hold substantial merits for Saturn's rings study as they enable us to conduct long-term observations, 
as well as to observe Saturn's rings at very low phase angles and low solar elevations, which are not covered by Cassini/CIRS.
In this paper, we show new high spatial resolution images of Saturn's rings in the MIR obtained with the 8.2-m Subaru Telescope in 2008 and 2005. We also discuss interesting features of the MIR emission from Saturn's rings including the temporal variation in the radial brightness contrast, and suggest possible causes of the variation.


\section{Data and reduction}
\subsection{Data set of observations in 2008}

Saturn and its rings were observed on January 23, 2008~(UT) 
with the COoled Mid-Infrared Camera and Spectrometer \citep[COMICS;][]{kataza00,okamoto02,sako03} mounted on Subaru Telescope (Program ID: S07B-076, PI: P.A.\,Yanamandra-Fisher). 
Imaging observations using the $8.8~\micron$ (the bandwidth $\Delta \lambda = 0.8~\micron$), $9.7~\micron$ ($\Delta \lambda = 0.9~\micron$), 
$10.5~\micron$ ($\Delta \lambda = 1.0~\micron$), $11.7~\micron$ ($\Delta \lambda = 1.0~\micron$), 
$12.5~\micron$ ($\Delta \lambda = 1.2~\micron$), $17.7~\micron$ ($\Delta \lambda = 0.9~\micron$), 
$18.8~\micron$ ($\Delta \lambda = 0.9~\micron$), $20.5~\micron$ ($\Delta \lambda = 0.9~\micron$), 
and $24.5~\micron$ ($\Delta \lambda = 0.8~\micron$) bands were carried out. 

We collected the data from the SMOKA data archive at National Astronomical Observatory of Japan (NAOJ). 
The pixel scale of the COMICS detector in the imaging mode is $0\farcs13$ pixel$^{-1}$.
To cancel out the background radiation, the secondary mirror was chopped at a frequency of $\sim 0.5$~Hz with a throw of 30$\arcsec$.
We also obtained imaging data of two MIR standard stars, HD~94336 \citep{cohen99} for the $N$-band (8--13$~\micron$) and $\alpha$~Boo \citep{cohen95} for the $Q$-band  (17--25$~\micron$), taken in the same manner on the same night close to the Saturn observations for flux calibration and point spread functions (PFSs). 
The observations are summarized in Table~\ref{obs2008}.
The astrometrical parameters of Saturn at the epoch of the observations are summarized in Table~\ref{parameter}.

It is noted that Saturn was observed at $7.8~\micron$ as well in the program, but the signal-to-noise ratio (SNR) in Saturn's rings was rather low because of the poor atmospheric transparency and faint brightness of the rings at the wavelength. Therefore we do not use the $7.8~\micron$ data in this study.

\begin{table}
\caption{Summary of observations on January 23, 2008 and April 30, 2005 (UT).\label{obs2008}}
\centering                          
\begin{tabular}{lccccc}
\hline\hline                 
Object & Filter      & Time  & Integ. & Airmass & FWHM\\
       & ($\micron$) & (UT)         & (s)         &   & ($\arcsec$)\\
\hline                        
\multicolumn{6}{l}{January 23, 2008 (UT): } \\
Saturn    & 8.8      & 12:16:27    & 242.2       & 1.02--1.03   & -- \\
HD~94336\tablefootmark{*}  & 8.8      & 14:28:27    & 10.8        & 1.06   & 0.54 \\
Saturn    & 9.7      & 12:35:45    & 240.6       & 1.01--1.02   & -- \\
HD~94336\tablefootmark{*}  & 9.7      & 14:32:27    & 10.3        & 1.07   & 0.58 \\
Saturn    & 10.5      & 13:01:04    & 181.3       & 1.01--1.02  & -- \\
HD~94336\tablefootmark{*}  & 10.5      & 14:34:33    & 10.6        & 1.07  & 0.48 \\
Saturn    & 11.7      & 11:47:56    & 42.4        & 1.05     & -- \\
HD~94336\tablefootmark{*}  & 11.7      & 14:36:53    & 10.6        & 1.08  & 0.38 \\
Saturn    & 12.5      & 13:17:32    & 123.4       & 1.02     & -- \\
HD~94336\tablefootmark{*}  & 12.5      & 14:38:53    & 10.6        & 1.08  & 0.49 \\
Saturn    & 17.7      & 13:37:28    & 30.3        & 1.03--1.04  & -- \\
$\alpha$~Boo\tablefootmark{*} & 17.7      & 15:04:06    & 10.1        & 1.06  & 0.52 \\
Saturn    & 18.8      & 13:48:06    & 30.6        & 1.04--1.05  & -- \\
$\alpha$~Boo\tablefootmark{*} & 18.8      & 15:07:09    & 10.2        & 1.06  & 0.56 \\
Saturn    & 20.5      & 13:57:20    & 40.8        & 1.05--1.06  & -- \\
$\alpha$~Boo\tablefootmark{*} & 20.5      & 15:09:41    & 10.2        & 1.06  & 0.57 \\
Saturn    & 24.5      & 14:09:09    & 42.4        & 1.07--1.08  & -- \\
$\alpha$~Boo\tablefootmark{*} & 24.5      & 15:12:09    & 10.6        & 1.05  & 0.67 \\
\hline                        
\multicolumn{6}{l}{April 30, 2005 (UT): } \\
Saturn    & 12.5      & 07:34:49    & 60.8       & 1.91-1.94    & -- \\
HD~120477\tablefootmark{*}  & 12.5      & 08:21:35    & 10.6        & 1.06  & 0.35 \\
Saturn    & 24.5      & 06:43:49    & 61.7        & 1.46--1.49  & -- \\
$\alpha$~Boo\tablefootmark{*} & 24.5      & 10:09:43    & 10.6        & 1.00  & 0.62 \\
\hline                                   
\end{tabular}
\tablefoot{\tablefoottext{*}{Standard star.}}
\end{table}

\begin{table}
\caption{Astrometrical parameters of observations with Subaru/COMICS in 2008 and 2005. \label{parameter}}
\centering                          
\begin{tabular}{cccccc}
\hline\hline                 
Epoch & $r$ & $\Delta$ & $B'$ & $B$ & $\alpha$ \\
(UT)  & (au) & (au) & (deg) & (deg) & (deg) \\
\hline                        
January 23, 2008 & 9.27 & 8.45 & $-8.7$ & $-7.2$ & 3.5 \\
April 30, 2005 & 9.07 & 9.33 & $-21.9 $ & $-23.6$ & 6.1 \\
\hline                                   
\end{tabular}
\tablefoot{$r$ is the heliocentric distance of Saturn, $\Delta$ the distance to the observer, 
$B'$ the elevation of the sun above the ring plane, $B$ the elevation of the observer above the ring plane, $\alpha$ the phase angle. $B'$ and $B$ are negative because the sun and observer are on the south of the rings. }
\end{table}

\subsection{Data reduction}

For data reduction, we used our own reduction tools and IRAF\footnote{IRAF is distributed by the National Optical Astronomy Observatory, which is operated 
by the Association of Universities for Research in Astronomy under cooperative agreement 
with the National Science Foundation.}. 
The standard chopping pair subtraction was employed. 
Flat fielding was achieved using sky frames taken immediately after the observations of Saturn. 
The shift-and-add method was applied for the flatfielded images 
and the stacked (median) image of Saturn was produced for each band. 

For flux calibration, a conversion factor (from ADU to flux density) in each band was calculated 
using the reduced images the standard stars and the photospheric models by \citet{cohen95,cohen99}.
We also measured a full-width at half-maximum (FWHM) of the standard star in each band (see Table~\ref{obs2008}).
Since the difference in the beam size is not negligible, 
we convolved the reduced Saturn images to a common spatial resolution (to that of the $24.5~\micron$ image) 
so that the final images could more directly be compared with one another.

\subsection{Supplementary data obtained in 2005}
 
Imaging observations of Saturn and its rings were also carried out with COMICS on April 30, 2005 (UT) (Program ID: S05A-029, PI: P.A.\,Yanamandra-Fisher) using the same filter set as that employed in 2008. 
However, the elevation of Saturn was much lower and the integration time was shorter for some filters than the observations in 2008, and the flat data were missing, all of which affect the image quality.  
In addition, standard stars were observed only in the $N$-, 12.5, and $Q$-, $24.5~\micron$ bands with a large discrepancy in airmass, resulting in the unsatisfactory absolute flux calibration. Due to the circumstances, 
we used only $12.5~\micron$ and $24.5~\micron$ images, which exhibited reasonable SNRs, 
as supplementary data to compare with those taken in 2008.
The observations are summarized in Table~\ref{obs2008}.
The astrometrical parameters of Saturn at the epoch of the observations are summarized in Table~\ref{parameter}.

The data reduction procedure was basically the same as for the 2008 data, except for the lack of flatfielding. At the flux calibration step, we applied an airmass correction by estimating the difference in atmospheric extinction between Saturn and the standard stars using ATRAN \citep{lord92}.

\section{Results}

\subsection{Saturn's rings observed in 2008}

\subsubsection{Images}

\begin{figure*}
\centering
\includegraphics[width=18cm]{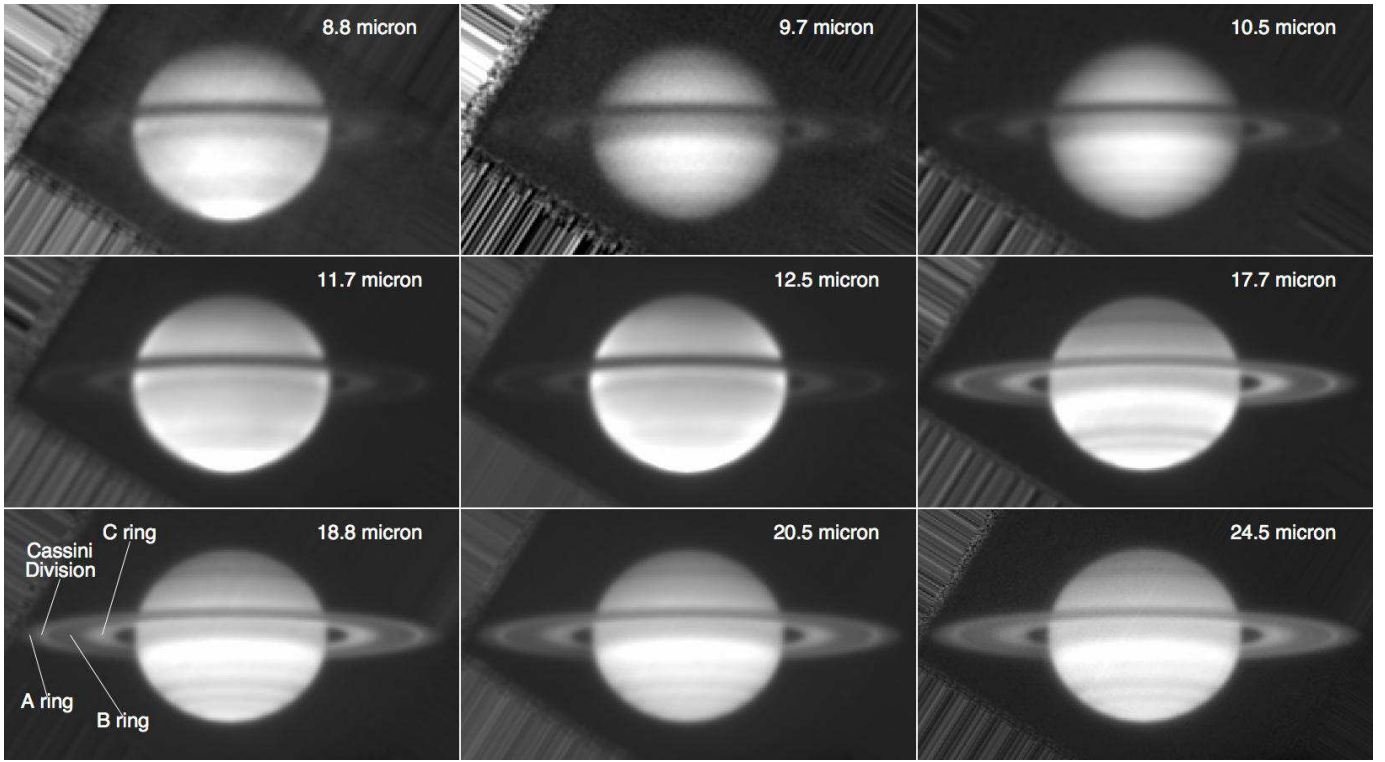}
\caption{MIR images of Saturn's rings on January 23, 2008 (UT) as seen by COMICS mounted on Subaru Telescope. 8.8~$\micron$ (top left), 9.7~$\micron$ (top middle), 10.5~$\micron$ (top right), 11.7~$\micron$ (middle left), 12.5~$\micron$ (middle middle), 17.7~$\micron$ (middle right), 18.8~$\micron$ (bottom left, shown with annotations of the rings' names), 20.5~$\micron$ (bottom middle), 24.5~$\micron$ (bottom right), all of which are shown in linear scale. North is up, and east is to the left. The field of view is 47\arcsec $\times$ 26\arcsec.
\label{image2008}}
\end{figure*}

The processed images of Saturn in 2008 at 8.8, 9.7, 10.5, 11.7, 12.5, 17.7, 18.8, 20.5, and $24.5~\micron$ are shown in Fig.~\ref{image2008}. The thermal emission from the rings is clearly seen in the images, in addition to the emission from the planetary body that is brighter than the rings. The contrast of the MIR brightness of the rings to the planetary body is smaller at the longer wavelength.

An interesting feature of the rings in the MIR brightness is two bright rings that are separated by a wide, fainter region that are seen in all the bands. The inner and outer bright rings are at the distance of 76000--86400 km (12\farcs5--14\farcs1) and 117000--121000 km (19\farcs2--19\farcs7) from the planet center, corresponding to the C ring and the Cassini Division, respectively. The detection of bright thermal emission from the Cassini Division suggests the presence a certain amount of material in the Cassini Division. The fainter region between those two bright rings corresponds to the B ring, which is the brightest in the visible light. Another faint ring just outside the Cassini Division is the A ring. The overall contrast of the MIR brightness of Saturn's rings observed in 2008 appears to be opposite to that in the visible light\footnote{A visible image of Saturn taken on March 16, 2008 (Japan Standard Time) by the 105-cm Murikabushi Telescope at NAOJ Ishigakijima Observatory in Okinawa, Japan is available at http://www.miz.nao.ac.jp/ishigaki/system/files/Sat\_IAO\_080316.jpg.}. The B ring looks like a shadow in the bright background of the planetary body, particularly in the $N$-band. It suggests that the B ring is opaque in the MIR.

\subsubsection{Profile of MIR brightness and color}

\begin{figure}
\centering
\includegraphics[width=9cm]{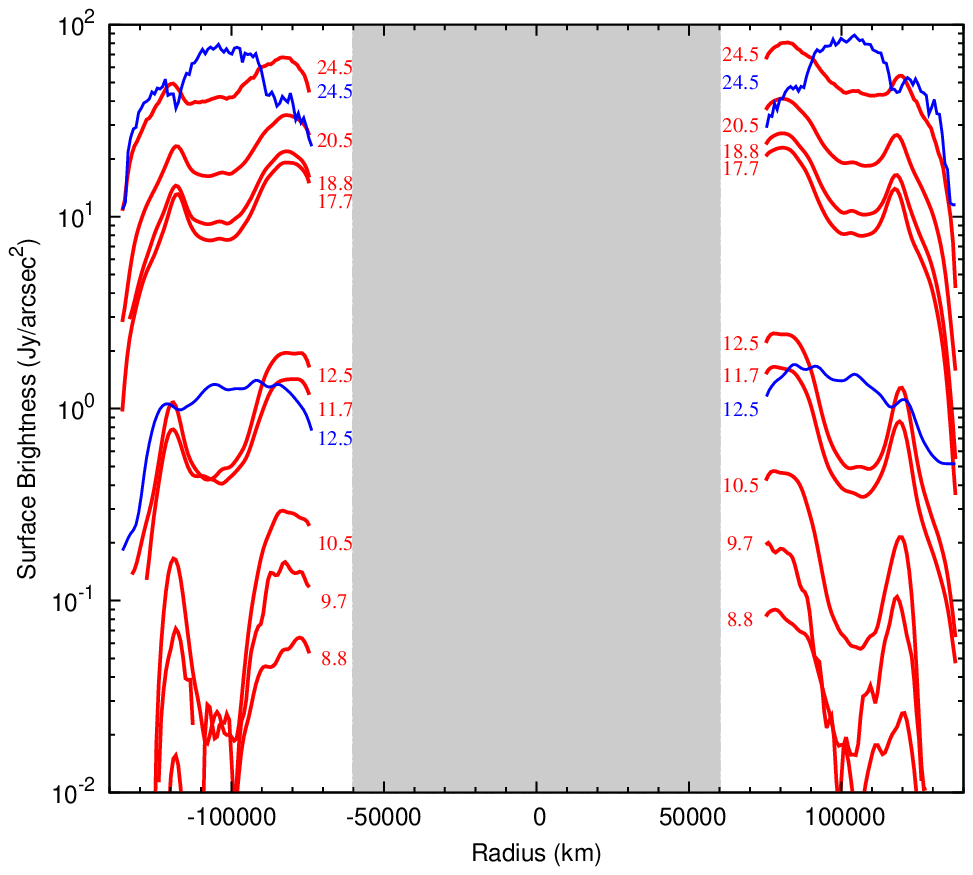}
\includegraphics[width=9cm]{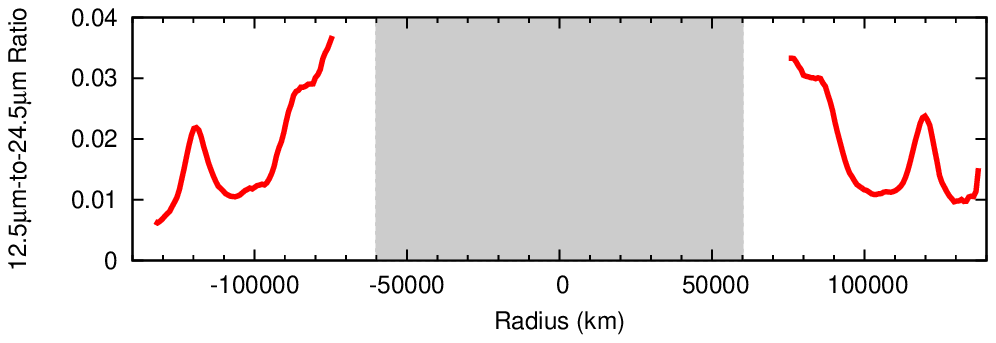}
\caption{Top: Profiles of surface brightness in all the bands in 2008 (red) and 2005 (blue). Bottom: Profile of 12.5$~\micron$-to-24.5$~\micron$ brightness ratio for Saturn's rings in 2008. Negative and positive value of the radius is for the east and west side, respectively. Region of the planetary body is shown in grey.
\label{profile}}
\end{figure}

To investigate the radial variation of the MIR emission from Saturn's rings, we extracted the averaged surface brightness in a width of 4 pixels (0\farcs52, slightly narrower than the beam size in the $24.5~\micron$ image) along the major axis of the projected shape of Saturn's rings. The derived profiles of the surface brightness in the MIR are shown in the upper panel of Fig.~\ref{profile}. The profiles in all the bands share a similar trend. There is a bump in the innermost region ($r=70000$--90000~km from the planet center) of the ring, corresponding to the C ring. Another bump is seen in the outer region around $r=120000$~km, which corresponds to the Cassini Division. Those two bumps are separated by a dip at $r=90000$--110000~km, which corresponds to the B ring. Fainter emission at $r \gtrsim 120000$~km, which corresponds to the A ring, is also detected, particularly at longer wavelengths. The contrast of the bumps to other regions is smaller at the longer wavelengths, which is probably because the B and A rings are cooler than the C ring and the Cassini Division. 

We also calculated $12.5~\micron$-to-$24.5~\micron$ color (brightness ratio) of the ring profile along the major axis of the projected shape of Saturn's rings. The resultant profile of the $12.5~\micron$-to-$24.5~\micron$ color of Saturn's rings is shown in the lower panel of Fig.~\ref{profile}. The color of the C ring and the Cassini Division is bluer than the B and A rings, suggesting that the C ring and the Cassini Division is warmer than the B and A rings.

\subsubsection{MIR SED of Saturn's rings}

\begin{figure}
\centering
\includegraphics[width=9cm]{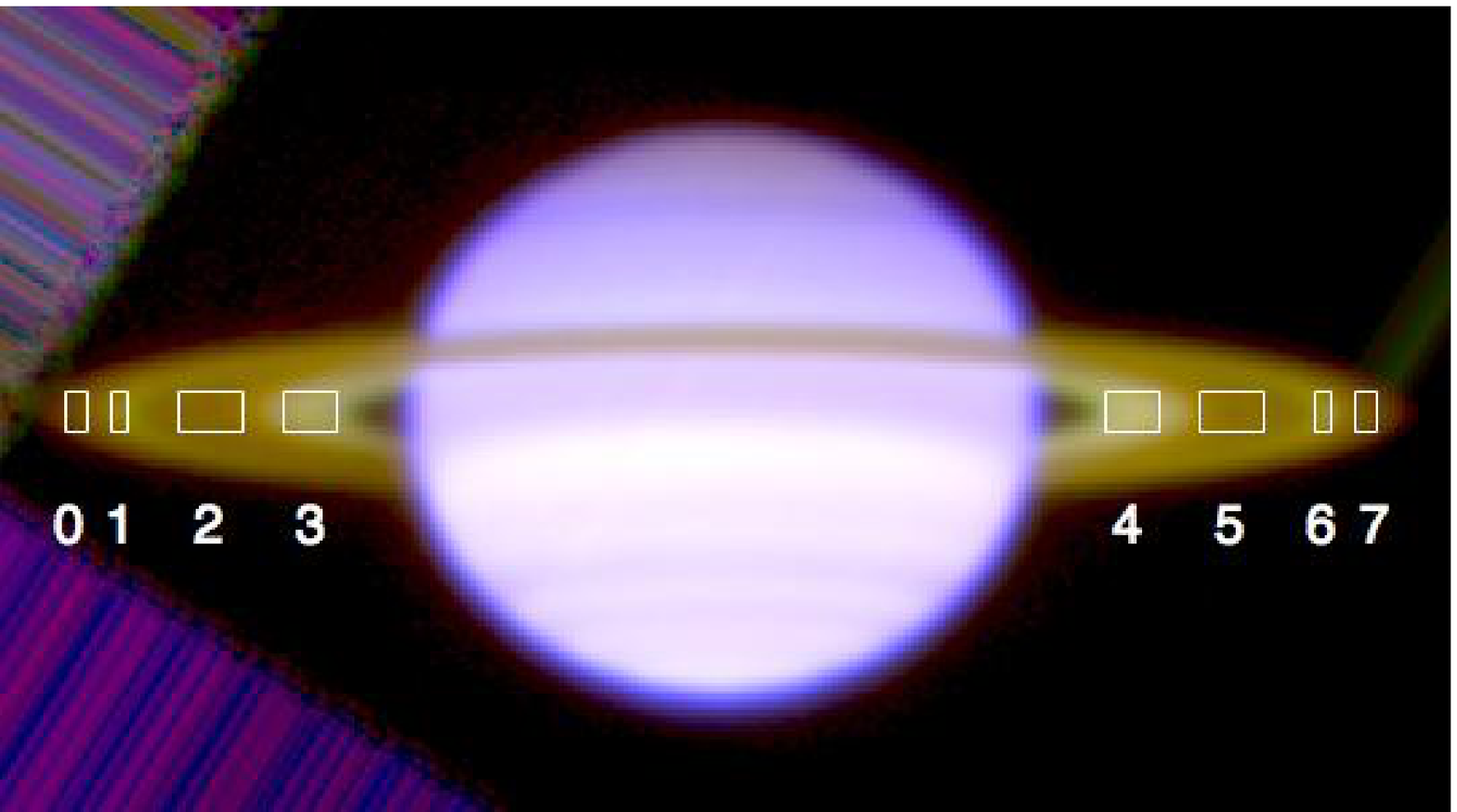}
\caption{3-band composite image of Saturn's rings using 12.5, 18.8, and 24.5$~\micron$. The white boxes indicate Regions 0--7 (from left to right). The field of view is 47\arcsec $\times$ 26\arcsec.
\label{region}}
\end{figure}

\begin{figure*}
\centering
\includegraphics[width=4.5cm]{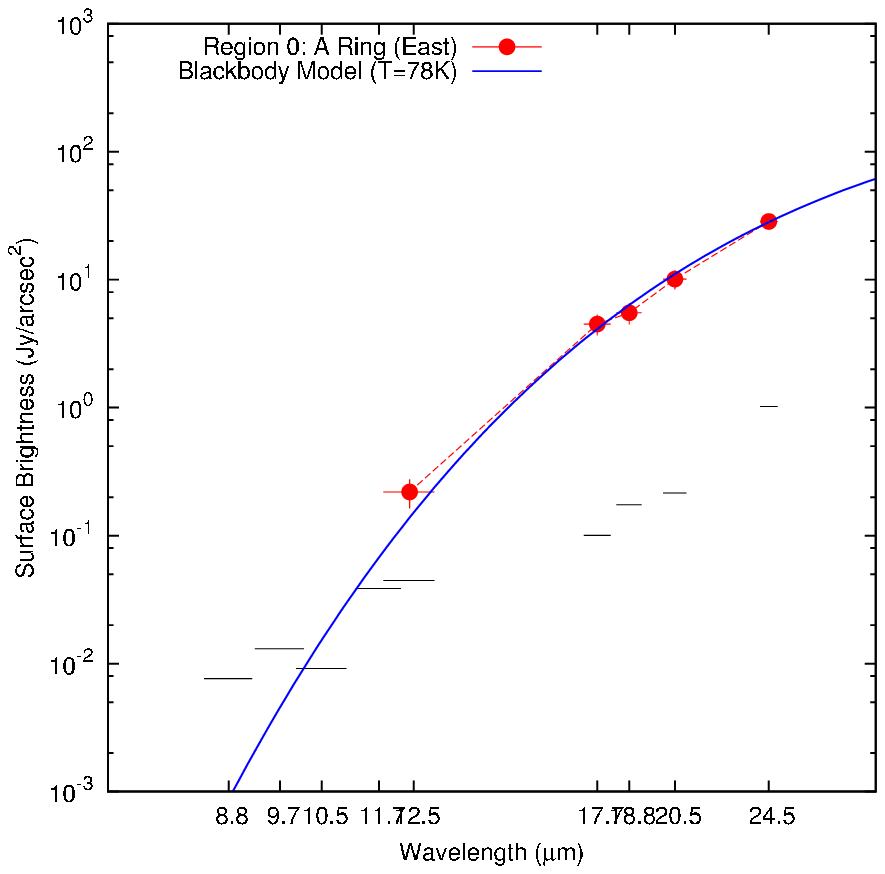}
\includegraphics[width=4.5cm]{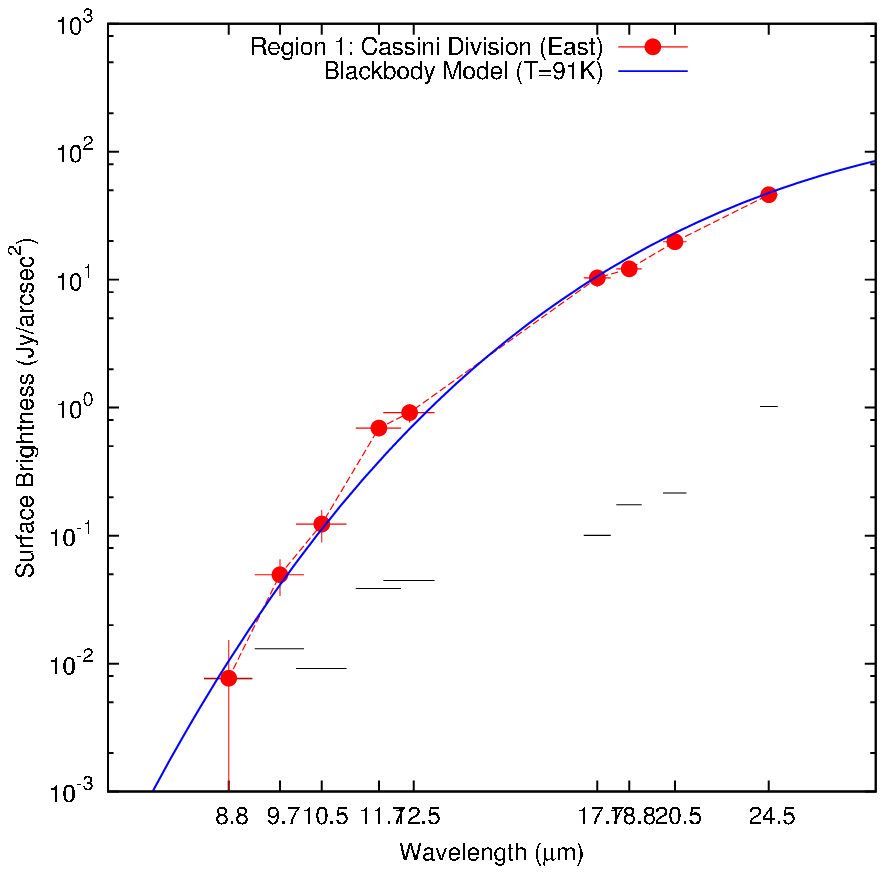}
\includegraphics[width=4.5cm]{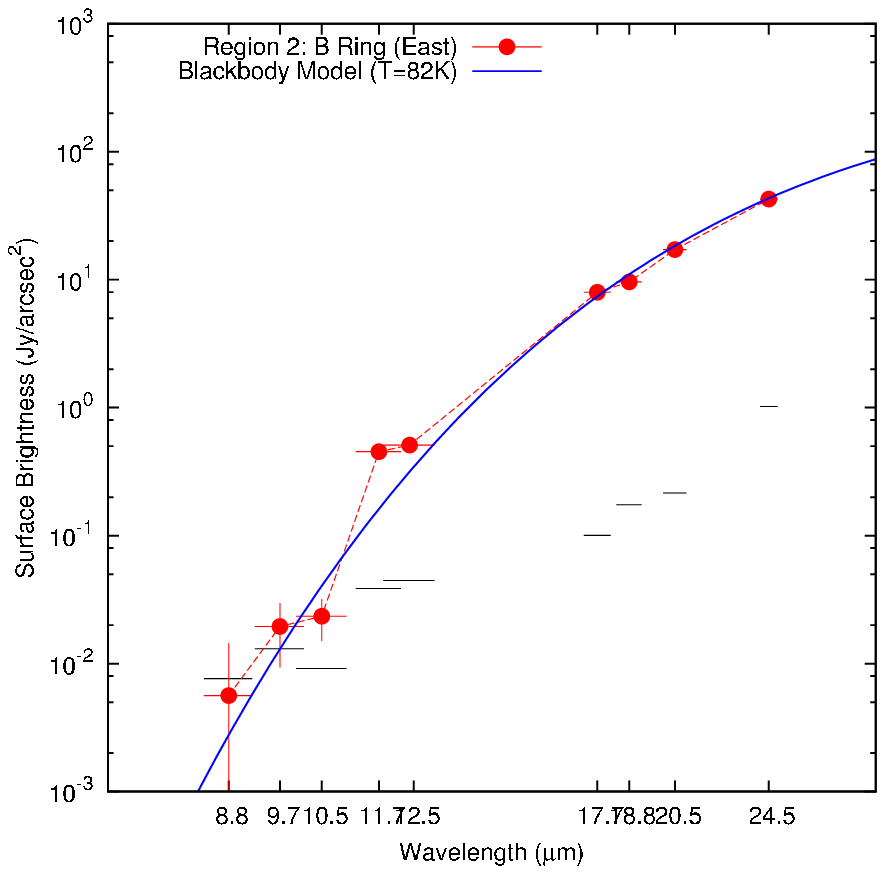}
\includegraphics[width=4.5cm]{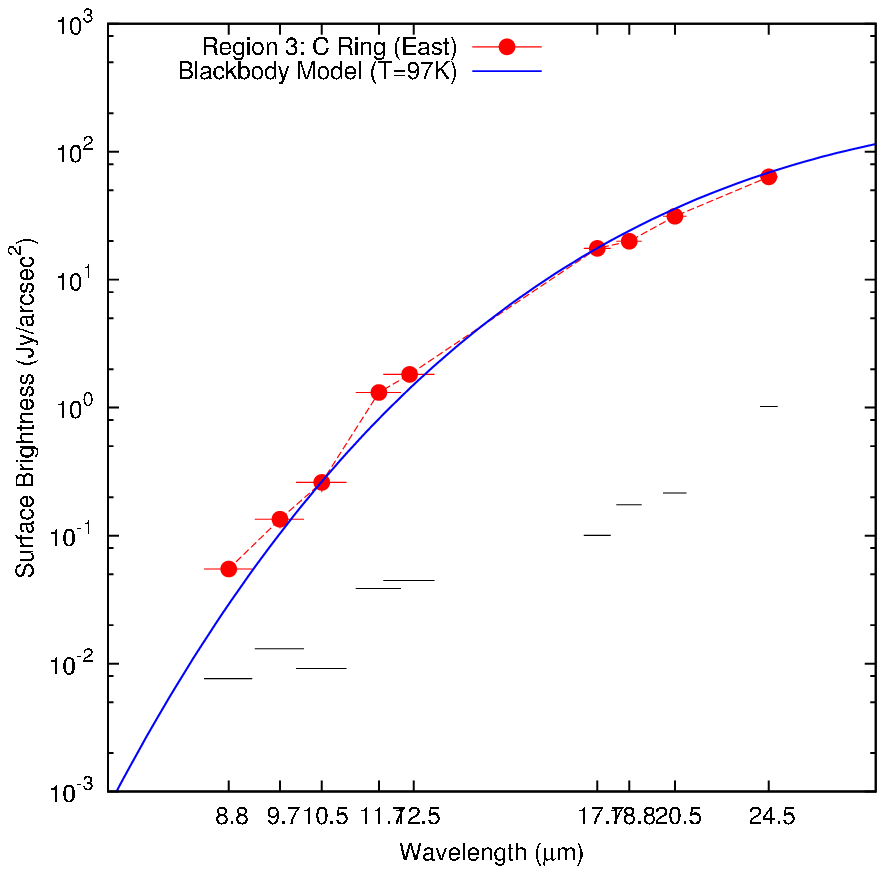}
\includegraphics[width=4.5cm]{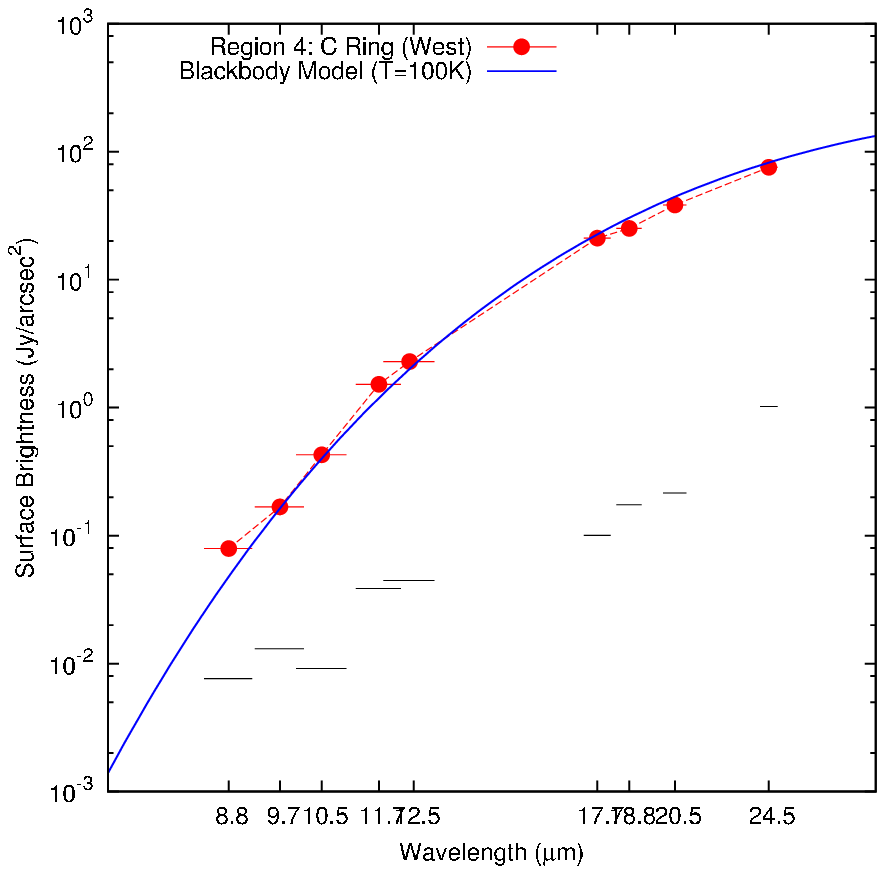}
\includegraphics[width=4.5cm]{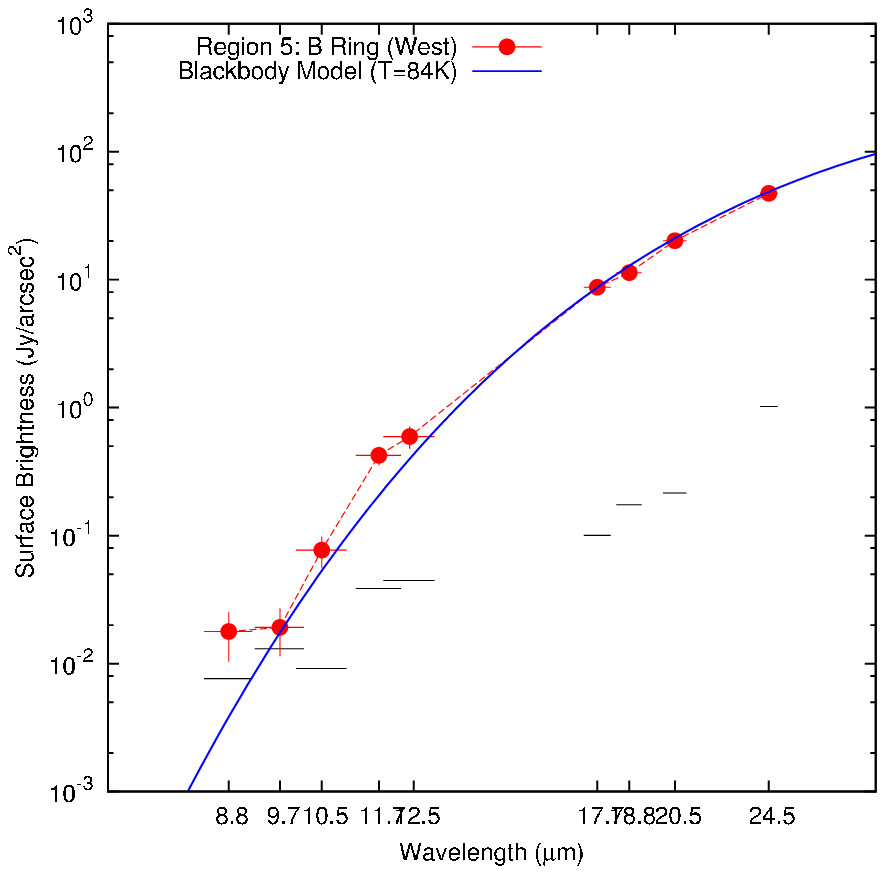}
\includegraphics[width=4.5cm]{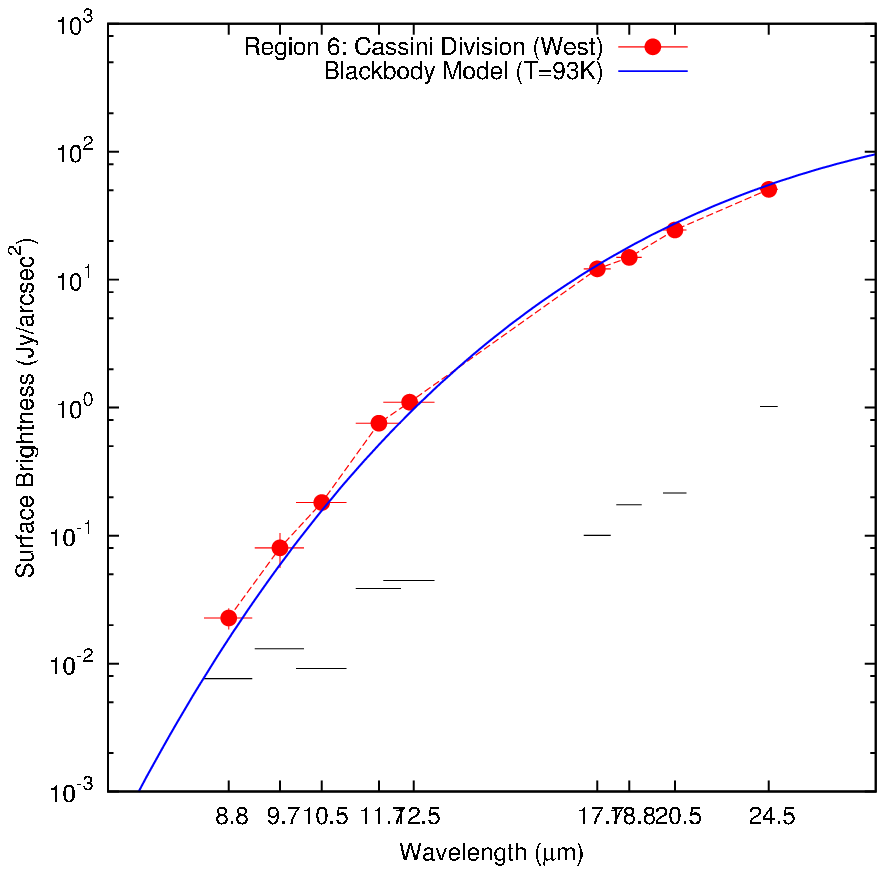}
\includegraphics[width=4.5cm]{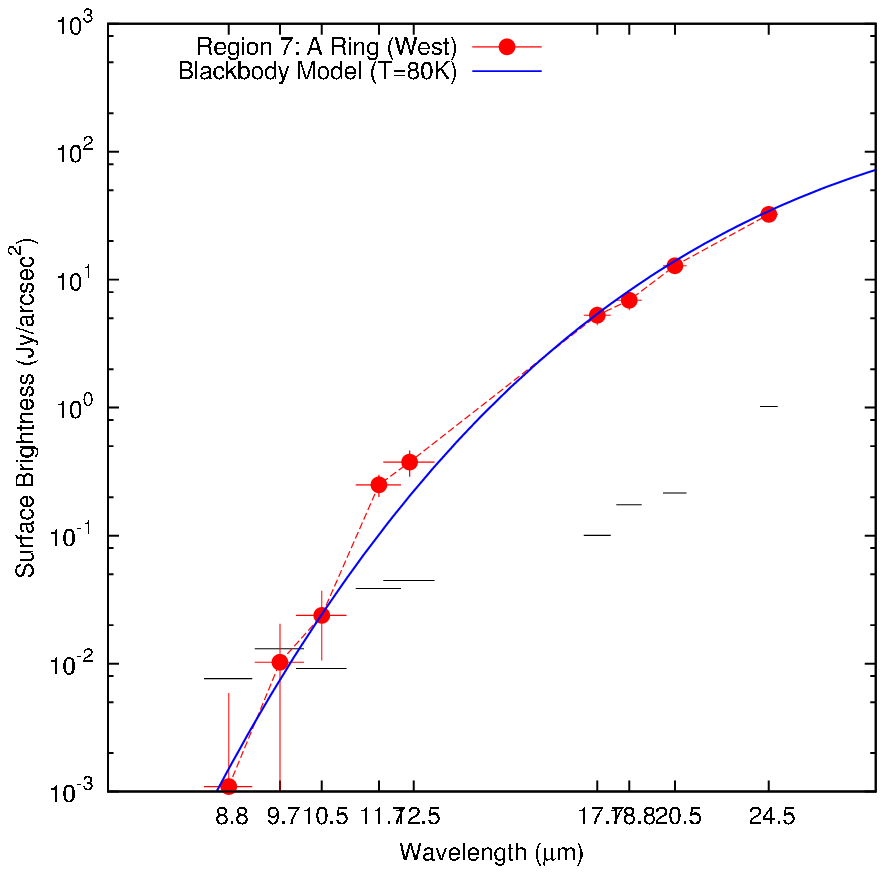}
\caption{MIR SEDs of Regions 0--7. Red points show the observations. Blue solid lines show the models of a single-temperature blackbody with the assumed optical depths of $\tau = 0.1$ for the C ring and the Cassini Division, $\tau = 0.5$ for the A ring, and $\tau = 1$ for the B ring. Horizontal black thin lines indicate the estimated standard deviations of background in the images.
\label{sed}}
\end{figure*}

To investigate the characteristics of Saturn's rings such as the brightness and color in the MIR and temperature quantitatively, we composed SEDs of the surface brightness of the thermal emission from selected regions in Saturn's rings based on the images taken by COMICS. The brightnesses of the detected emission were calibrated reasonably well by observing the flux calibrators \citep{cohen99} in all the bands. Eight regions, which correspond to the east and west sides of the C, B, and A rings and the Cassini Division, were defined as shown in Fig.~\ref{region}. We measured the averaged surface brightness in each region in each band (Table~\ref{photometry}), and composed the SEDs (Fig.~\ref{sed}). It is noted that the region 0 (the east ansa of the A ring) is partly out of the field of view and and no reliable measurement is available at 8.8, 9.7, 10.5 and 11.7$~\micron$. 
We also estimated standard deviations of the background in a blank region at each band (shown in Table~\ref{photometry} and Fig.~\ref{sed}), and treated them as photometric errors. 
The SEDs from 8.8 to $24.5~\micron$ all show a blackbody-like smooth shape with a rising surface brightness as the wavelength increases. The peak of the emission from Saturn's rings is located at $\lambda > 25~\micron$, suggesting that the rings are cooler than 200~K.

The SEDs of the rings measured in the COMICS images, $I_\nu(\lambda)$, can be modeled simply by 
\begin{equation}
I_\nu(\lambda) = \beta \, B_\nu(\lambda, T), 
\label{sedmodel}
\end{equation}
where $\beta$ is the filling factor of particles in the rings and $B_\nu(\lambda, T)$ is the Planck function of the temperature $T$. 
The filling factor is expressed analytically by
\begin{equation}
\beta = 1 - \exp \frac{- \tau}{|\sin B|}, 
\label{betamodel}
\end{equation}
where $\tau$ is the normal optical depth of the rings and $B$ the elevation of the observer above the ring plane.  

The optical depths of Saturn's rings have been measured from stellar occultations. Stellar occultation observations by the Ultraviolet Imaging Spectrograph Subsystem (UVIS) onboard Cassini provide the profiles of the optical depths in Saturn's rings at a spatial resolution of 10~km or higher. The UVIS measurements indicate the optical depths for both the C ring and the Cassini Division average at $\tau \sim 0.1$ partly ranging from $\sim 0.05$ to $\sim 0.2$, while that of the B and A rings is $\tau \gtrsim 1$ and $\tau \gtrsim 0.5$, respectively \citep{colwell09,colwell10}. 

For each region, we derived the ring temperature $T$ as the value that minimizes the weighted residual between the measured SED and a model spectrum defined by Eq.~(\ref{sedmodel}), assuming the optical depths of $\tau=0.1$ for the C ring and the Cassini Division, $\tau=1$ for the B ring, and $\tau=0.5$ for the A ring. The photometric errors (i.e., the background standard deviations) were fully taken into account in the residual minimization process.
We also calculated $T$ by changing $\tau$ by a factor of 2 to see the dependency on the assumed optical depth.
Temperatures derived with the assumed optical depth of each region are listed in Table~\ref{T}. 
The ring temperature estimated from the COMICS images is $T=97$--$100$~K for the C ring (Regions 3 and 4), $T=82$--$84$~K for the B ring (Regions 2 and 5), $T=78$--$80$~K for the A ring (Regions 0 and 7), and $T=91$--$93$~K for the Cassini Division (Regions 1 and 6). 
The uncertainty in the derived $T$ in the SED fitting is $\lesssim1$~K for all the regions. 
A considerable radial variation in temperature of the rings is found---the C ring is the warmest, the Cassini Division second, the B ring third, and the A ring fourth among the investigated regions. 
We note that the derived temperature of the narrow Cassini Division might be underestimated due to possible photometric contamination from the adjoining B and A rings. 
However the overall contrast in the temperatures of C, B, A rings and the Cassini Division presented here is secure. 

The best-fit SEDs of the eight regions are shown in Fig.~\ref{sed}. 
As can be seen, the SEDs observed in all the regions can be represented reasonably well by a single-temperature blackbody. It is noted that all the SEDs exhibit a hint of a small bump over the continuum at around 11.7~$\micron$ in all the regions. 
\cite{lynch00} presented featureless spectra of the A and B rings in the $N$-band (at $B=19.1^\circ$ and $B'=20.6^\circ$) taken at the 3-m Infrared Telescope Facility on June 16, 1991 (UT). 
Future $N$-band spectroscopic observations with higher accuracies at similar $B$ and $B'$ are required to confirm the possible feature in Saturn's rings.

\begin{table*}
\caption{Surface brightness of each region. The standard deviation of the background (BG) at each band is also shown. \label{photometry}}
\centering                          
\begin{tabular}{cccccccccc}
\hline\hline                 
Region & 8.8$~\micron$ & 9.7$~\micron$ & 10.5$~\micron$ & 11.7$~\micron$ & 12.5$~\micron$ & 17.7$~\micron$ & 18.8$~\micron$ & 20.5$~\micron$ & 24.5$~\micron$ \\
 & {\small (Jy/arcsec$^{2}$)} & {\small (Jy/arcsec$^{2}$)} & {\small (Jy/arcsec$^{2}$)} & {\small (Jy/arcsec$^{2}$)} & {\small (Jy/arcsec$^{2}$)} & {\small (Jy/arcsec$^{2}$)} & {\small (Jy/arcsec$^{2}$)} & {\small (Jy/arcsec$^{2}$)} & {\small (Jy/arcsec$^{2}$)} \\
\hline                        
0 & -- & -- & -- & -- & $2.2 \times 10^{-1}$ & $4.5 \times 10^{+0}$ & $5.5 \times 10^{+0}$ & $1.0 \times 10^{+1}$ & $2.8\times 10^{+1}$ \\ 
1 & $7.7 \times 10^{-3}$ & $5.0 \times 10^{-2}$ & $1.2 \times 10^{-1}$ & $6.9 \times 10^{-1}$ & $9.1 \times 10^{-1}$ & $1.0 \times 10^{+1}$ & $1.2 \times 10^{+1}$ & $2.0 \times 10^{+1}$ & $4.6\times 10^{+1}$ \\ 
2 & $5.6 \times 10^{-3}$ & $1.9 \times 10^{-2}$ & $2.3 \times 10^{-2}$ & $4.5 \times 10^{-1}$ & $5.1 \times 10^{-1}$ & $8.0 \times 10^{+0}$ & $9.6 \times 10^{+0}$ & $1.7 \times 10^{+1}$ & $4.3\times 10^{+1}$ \\ 
3 & $5.5 \times 10^{-2}$ & $1.3 \times 10^{-1}$ & $2.6 \times 10^{-1}$ & $1.3 \times 10^{+0}$ & $1.8 \times 10^{+0}$ & $1.8 \times 10^{+1}$ & $2.0 \times 10^{+1}$ & $3.1 \times 10^{+1}$ & $6.4\times 10^{+1}$ \\ 
4 & $7.9 \times 10^{-2}$ & $1.7 \times 10^{-1}$ & $4.3 \times 10^{-1}$ & $1.5 \times 10^{+0}$ & $2.3 \times 10^{+0}$ & $2.1 \times 10^{+1}$ & $2.5 \times 10^{+1}$ & $3.8 \times 10^{+1}$ & $7.6\times 10^{+1}$ \\ 
5 & $1.8 \times 10^{-3}$ & $1.9 \times 10^{-2}$ & $7.7 \times 10^{-2}$ & $4.2 \times 10^{-1}$ & $6.0 \times 10^{-1}$ & $8.7 \times 10^{+0}$ & $1.1 \times 10^{+1}$ & $2.0 \times 10^{+1}$ & $4.7\times 10^{+1}$ \\ 
6 & $2.3 \times 10^{-2}$ & $8.0 \times 10^{-2}$ & $1.8 \times 10^{-1}$ & $7.6 \times 10^{-1}$ & $1.1 \times 10^{+0}$ & $1.2 \times 10^{+1}$ & $1.5 \times 10^{+1}$ & $2.4 \times 10^{+1}$ & $5.1\times 10^{+1}$ \\ 
7 & $1.1 \times 10^{-3}$ & $1.0 \times 10^{-2}$ & $2.4 \times 10^{-2}$ & $2.5 \times 10^{-1}$ & $3.8 \times 10^{-1}$ & $5.3 \times 10^{+0}$ & $6.9 \times 10^{+0}$ & $1.3 \times 10^{+1}$ & $3.2\times 10^{+1}$ \\ 
\hline                        
BG & $7.6 \times 10^{-3}$ & $1.3 \times 10^{-2}$ & $9.3 \times 10^{-3}$ & $3.9 \times 10^{-2}$ & $4.5 \times 10^{-2}$ & $1.0 \times 10^{-1}$ & $1.7 \times 10^{-1}$ & $2.2 \times 10^{-1}$ & $1.0\times 10^{+0}$ \\ 
\hline                                   
\end{tabular}
\end{table*}

\begin{table*}
\caption{Temperatures of each region in 2008 derived from the COMICS-observed SED by assuming the optical depths. Boldfaces show the values based on the most likely optical depths measured from stellar occultation observations by Cassini/UVIS.  \label{T}}
\centering                          
\begin{tabular}{clcccccc}
\hline\hline                 
Region & Ring & \multicolumn{6}{c}{$T$ by COMICS (K)} \\
 \cline{3-8}
&  &  $\tau=0.05$ & $0.1$ & $0.2$ & $0.5$ & $1$ & $2$ \\
\hline                        
0 & A ring (East)  & -- & -- & 80 & {\bf 78} & 78 & -- \\
1 & Cassini Division (East) & 97 & {\bf 91} & 87 & -- & -- & -- \\
2 & B ring (East)  & -- & -- & -- & 82 & {\bf 82} & 82 \\
3 & C ring (East)  & 102 & {\bf 97} & 92 & -- & -- & -- \\
4 & C ring (West)  & 105 & {\bf 100} & 95 & -- & -- & -- \\
5 & B ring (West)  & -- & -- & -- & 84 & {\bf 84} & 84 \\
6 & Cassini Division (West) & 99 & {\bf 93} & 89 & -- & -- & --  \\
7 & A ring (West)  & -- & -- & 82 & {\bf 80} & 80 & -- \\
\hline                                   
\end{tabular}
\end{table*}

\subsection{Saturn's rings observed in 2005}

\begin{figure}
\centering
\includegraphics[width=9cm]{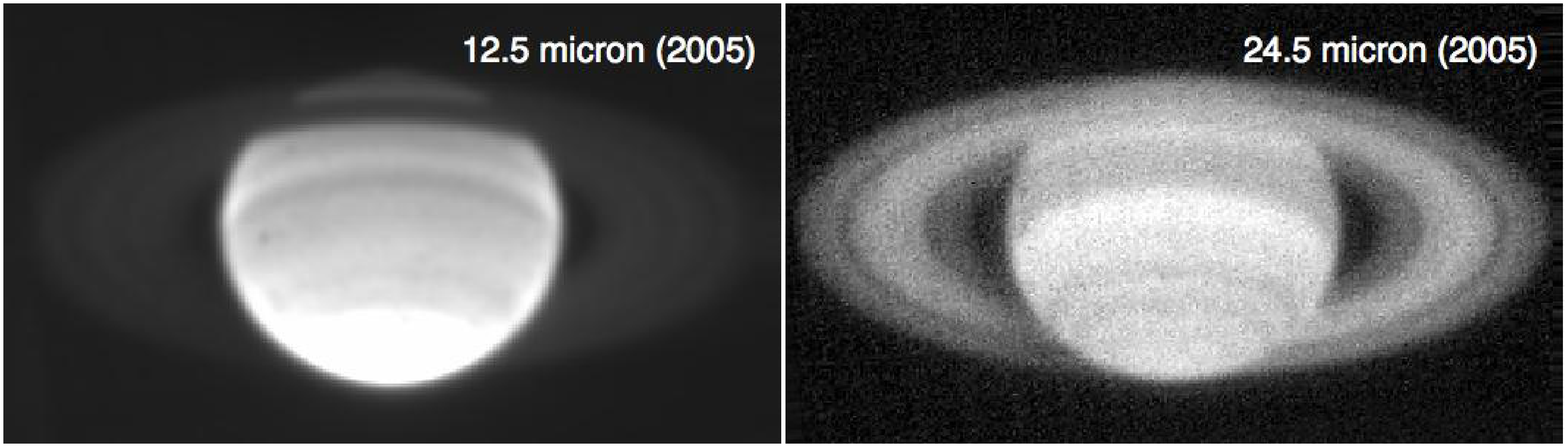}
\includegraphics[width=9cm]{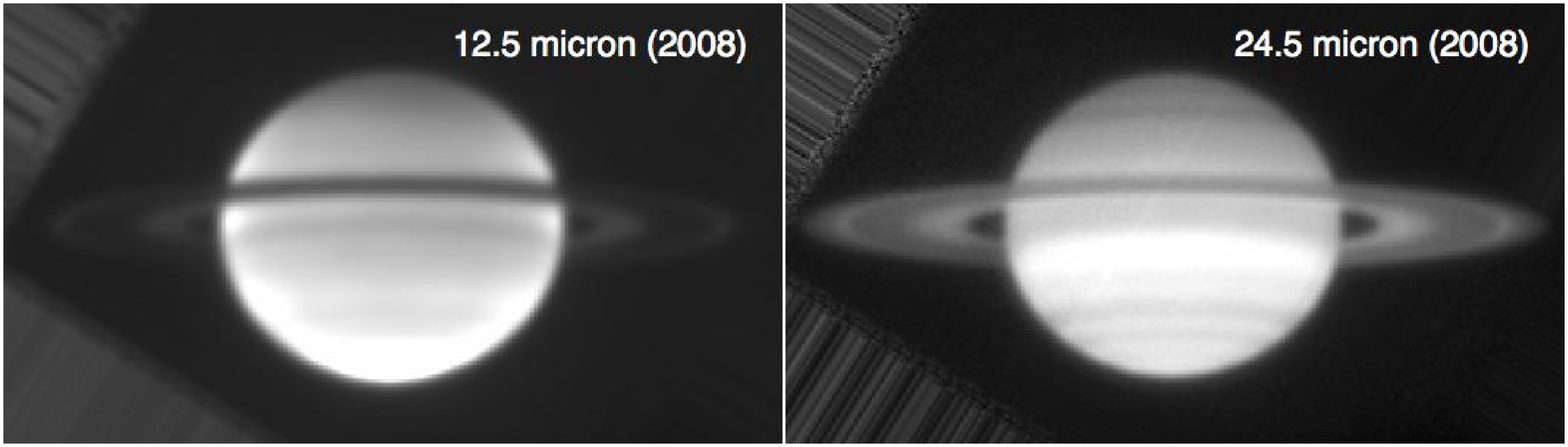}
\caption{Comparison of MIR images of Saturn's rings in 2005 (top) and 2008 (bottom). 12.5~$\micron$ (left) and 24.5~$\micron$ (right). The images in 2005 are scaled by the distance to the observer to show the same region as the 2008 images. The field of view of each panel is 47\arcsec $\times$ 26\arcsec.
\label{image2005}}
\end{figure}

The processed images of Saturn's rings in 2005 at 12.5 and $24.5~\micron$ are shown in the upper panels of Fig.~\ref{image2005} (those taken in 2008 are also displayed in the lower panels as comparisons). 
The radial profiles of the surface brightness are overplotted on the 2008 data in Fig~\ref{profile}.
The images are consistent with the ones presented by \citet{fletcher09}, who discussed the atmospheric features of Saturn itself using the same COMICS data set as this study. \citet{leyrat08b} also presented a image of Saturn in 2005 observed at 19.5$~\micron$ using VLT/VISIR. Their image showed similar characteristics to those captured at $24.5~\micron$ by COMICS. 

In contrast to 2008, 
the B ring is the brightest, the A rings the second, and the C ring and the Cassini Division is fainter, in the $24.5~\micron$ image taken in 2005. 
The overall radial brightness contrast of Saturn's rings at $24.5~\micron$ observed in 2005 is similar to the image in the visible light.

The rings in the $12.5~\micron$ image in 2005 seem relatively flat in surface brightness, 
which is also different from the 2008 image. In addition, a few narrow bright features can bee seen in the B and C rings. 
The B ring looks opaque, while the C ring and the Cassini Division are relatively transparent, when seen against the background of the planetary body.

\section{Discussion}

\subsection{Spatial variations in temperature}

We estimated the temperatures of the C, B, and A rings and the Cassini Division from the MIR images taken by COMICS in 2008 by assuming the optical depths. 
We found that the C ring was the warmest and the Cassini Division second, the B ring third, and the A ring fourth at the epoch of the observations in 2008. 

Higher temperatures of the C ring and the Cassini Division compared with the B and A rings are accounted for by their lower albedos, which might be a result of the pollution of ring material due to meteoroid bombardment \citep{cuzzi98}. 
At the same time, the lower optical depths of the C ring and the Cassini Division could also contribute to their higher temperatures since lower optical depths reduce mutual obscuration and shadowing between particles in the rings \citep{spilker06}.
We note that our measurements suggest that the temperature at the east (morning) ansa is slightly lower than that at the west (afternoon) ansa of each ring. This could be accounted for by eclipse cooling in Saturn's shadow.

\subsection{C ring and Cassini Division brighter than B and A rings in 2008}

One of the most interesting features in Saturn's rings seen in our COMICS images taken in 2008 is an inversion of the contrast of the MIR surface brightness in the C, B, and A rings and the Cassini Division, compared with the visible image. The C ring and the Cassini Division were brighter than the B and A rings in the MIR in 2008. 
Saturn was observed in the MIR in 2005 by Subaru/COMICS \citep[this work;][]{fletcher09} and VLT/VISIR \citep{leyrat08b}. 
The overall radial contrast of the MIR brightness in 2005 appears to be the inverse of that in 2008 and is similar to images in the visible light---there is a dip in the surface brightness at the position of the Cassini Division, and the C ring and the Cassini Division is fainter than the B and A rings in the $Q$-band. 

The surface brightness of a ring in the thermal emission is expressed as a product of the particle filling factor and the Planck function of the temperature (see Eq.~(\ref{sedmodel})). 
Our COMICS measurements at low $\alpha$ suggest the temperature contrast between optically thick, cooler rings (the B and A rings) and optically thin, warmer rings (the C ring and Cassini Division) is larger at a low $B'$ (or $B$) in 2008 than at a high $B'$ (or $B$) in 2005. 
On the other hand, the contrast in the particle filling factors seen from us is smaller in 2008 than in 2005. 
Both of the contrast changes in the ring temperatures and filling factors result in the brighter C and Cassini Division than the B and A rings in the MIR in 2008 as seen in our images.

To summarize, the inversion of the MIR contrast of Saturn's rings from 2005 to 2008, which made the C ring and the Cassini Division brighter than the B and A rings in 2008, is accounted for by a seasonal effect with changing elevations of the sun and observer above the ring plane. 
In 2008, higher temperatures of the C ring and the Cassini Division than the B and A rings, and the increased filling factors of particles in the C ring and the Cassini Division resulted in the higher MIR surface brightness of the C ring and the Cassini Division than the B and A rings. Although the temperatures of the C ring and the Cassini Division are higher than the B and A rings also in 2005, their low filling factors result in the relatively lower brightnesses compared with the B and A rings. 

\cite{froidevaux81} compiled a list of brightness temperatures of the A, B, and C rings measured at $20~\micron$ at various $B'$ from the literature by different groups and instruments. 
They suggested that the C ring is brighter than the B and A rings at $|B| \sim |B'|\lesssim 11^\circ$ \citep[see Table~1 in][]{froidevaux81}. 
Our images of Saturn's rings from the single instrument, COMICS, show new, clear visualized evidence of the higher MIR surface brightness of the Cassini Division as well as the C ring, than the B and A rings at a lower elevation of the sun and observer above the ring plane. 
The present work resolves the Cassini Division in the thermal emission by ground-based observations for the first time.

We note that, unlike in the thermal emission, seasonal reversal of brightness contrast over the rings does not occur in the visible light (for instance, see Fig.~1 in \cite{cuzzi02} for images of Saturn rings at various solar elevations observed by the Hubble Space Telescope). This is simply because the reflected brightnesses of the C ring and the Cassini Division, which have low albedo particles, are always lower than those for the B and A rings even after filling factor corrections.

\subsection{Comparison with Cassini/CIRS temperature} 

To make a direct comparison of the ring temperatures by COMICS with those by Cassini/CIRS, we computed the temperature profiles from the MIR brightnesses measured by COMICS (shown in Fig.~\ref{profile}). A radial profile of the normal optical depth in the rings (PRODUCT ID: UVIS\_HSP\_2006\_256\_LAMCET\_I\_TAU\_10KM.TAB) measured from stellar occultation observations by Cassini/UVIS\footnote{Data retrieved from The Ring-Moon Systems Node of NASA's Planetary Data System http://pds-rings.seti.org.} \citep{colwell09,colwell10} was converted to the particle filling factors at the epoch of COMICS observations in 2005 and 2008 by Eq.~(\ref{betamodel}), and then smoothed with the COMICS' PSFs at $24.5~\micron$. We performed fits with an SED model defined by Eq.~(\ref{sedmodel}) and derived the ring temperature at each radius.

\begin{figure*}
\centering
\includegraphics[width=9cm]{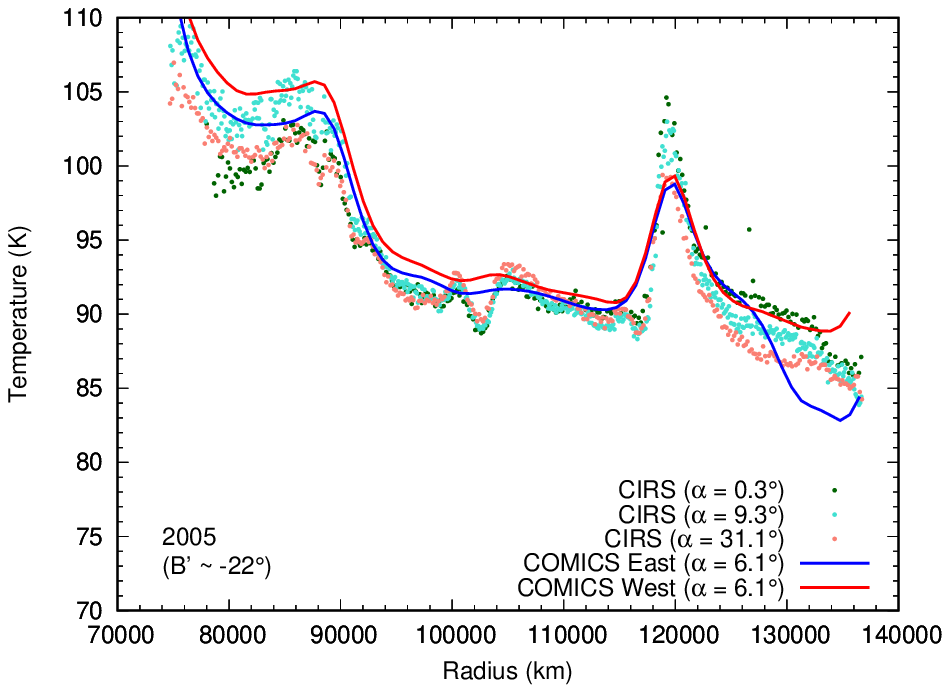}
\includegraphics[width=9cm]{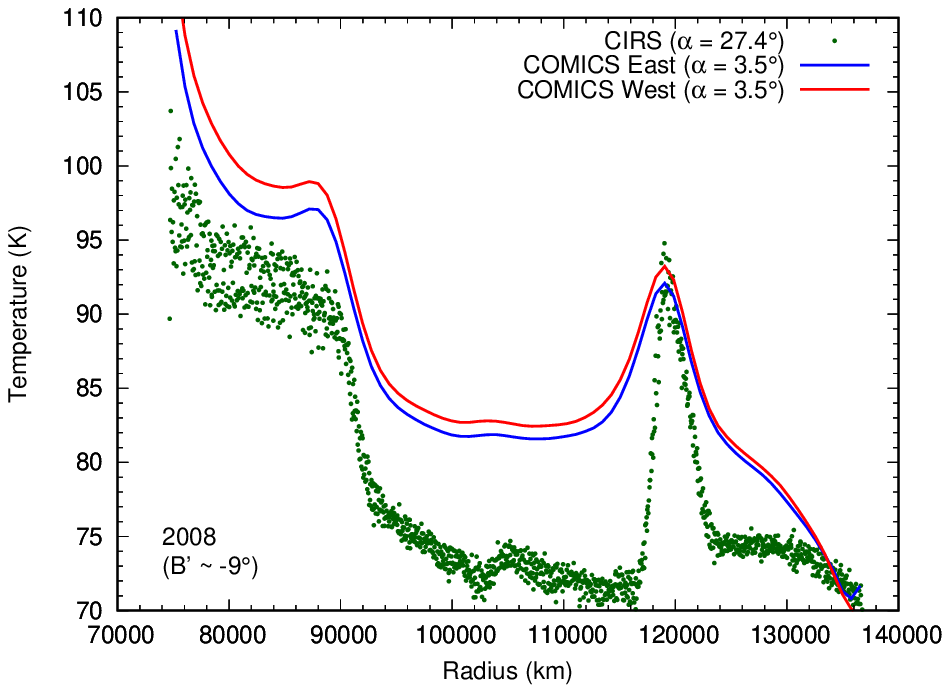}
\includegraphics[width=9cm]{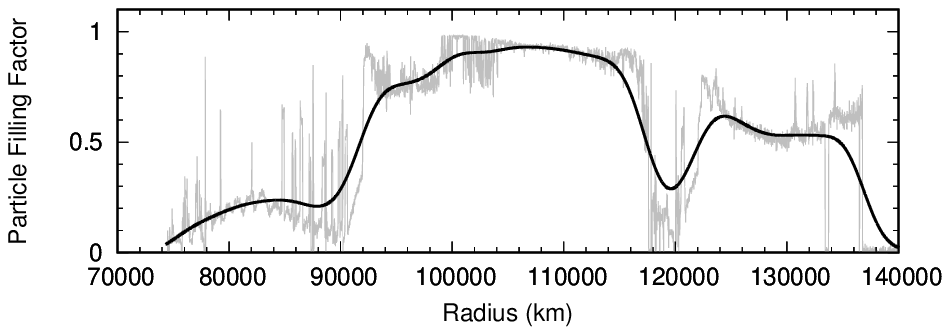}
\includegraphics[width=9cm]{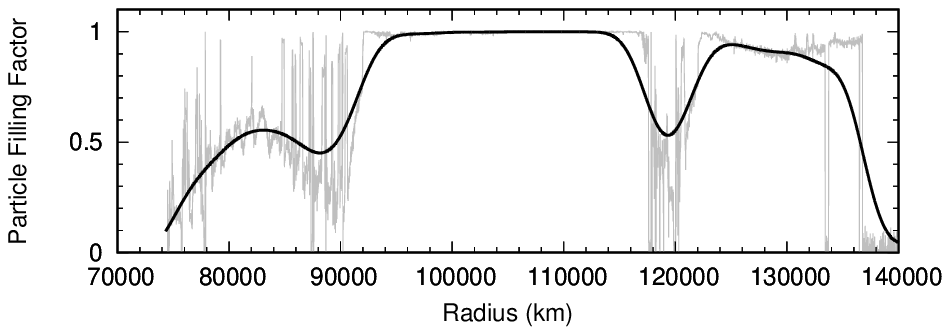}
\caption{Top: Profiles of ring temperature in 2005 (left) and 2008 (right), calculated from the COMICS-measured MIR SEDs and the particle filling factors based on the Cassini/UVIS-measured optical depths. 
Blue and red lines indicate the east (morning) and west (afternoon) ansae, respectively.  
Ring temperatures based on the Cassini/CIRS FP1 data are also overplotted. 
Bottom: Profiles of the particle filling factors with the radial resolution of 10~km of the original UVIS data products (gray lines) and those are smoothed with the beam sizes of COMICS (black thick lines). 
\label{temperature}}
\end{figure*}

The resultant profiles of the ring temperature are shown in Fig.~\ref{temperature}. 
The achieved spatial resolution of the profiles is $\sim~4000$~km, and, to our best knowledge, seems the highest ever reported from ground-based observations. 
The temperatures in 2008 are consistent with the values estimated for the selected eight regions shown in Table~\ref{T}. 
Both in 2005 and 2008, the east (morning) ansa is cooler than the west (afternoon) ansa over the rings, which could be accounted for by eclipse cooling in Saturn's shadow.

Cassini/CIRS has been making a huge number of observations of Saturn's rings since 2004. 
For comparison with our COMICS results, CIRS FP1 data of three radial scans in 2005 and one in 2008 with close observational epoch and solar elevation above the ring plane to the COMICS observations were collected. 
The astrometrical parameters of Saturn at the epoch of the CIRS observations are summarized in Table~\ref{CIRSparameter}.
The ring temperatures in each scan were derived from fits with a spectral model of blackbody \citep[e.g.][]{spilker06,altobelli08}.

In Fig.~\ref{temperature}, ring temperature profiles measured by CIRS are overplotted on the COMICS data. 
The profile measured by COMICS in 2005 is compared with those by CIRS in 2005 at the phase angle of $\alpha = 0.3$, $9.3$, $31.0^\circ$, and is in agreement with them. 
It is noted that the COMICS measurements at $\alpha = 6.1^\circ$ suggest slightly higher temperatures than the CIRS data at similar phase angles ($\alpha = 0.3$ and $9.3^\circ$), which might be due to the difference in the observation wavelength. 
MIR observations by COMICS should be more sensitive to warmer particle in the rings than FIR observations by CIRS FP1 channel. 
The compiled data by COMICS and CIRS in 2005 suggest small change in ring temperature profile between $\alpha \sim 0^\circ$ and $\sim 30^\circ$ at a large solar elevation above the ring plane ($B' \sim -22^\circ$).

\begin{table*}
\caption{Astrometrical parameters of observations with Cassini/CIRS in 2005 and 2008. \label{CIRSparameter}}
\centering                          
\begin{tabular}{ccccccc}
\hline\hline                 
Epoch & Ring Observation ID & $r$ & $\Delta$ & $B'$ & $B$ & $\alpha$ \\
(UT)  &           & (au) & (au) & (deg) & (deg) & (deg) \\
\hline                        
April 14, 2005 & S\_SPEC\_CO\_CIRS\_1492152751\_FP1 & 9.07 & 9.08 & $-22.0$ & $-7.6$ & 31.1 \\
May 20, 2005   & S\_SPEC\_CO\_CIRS\_1495265272\_FP1 & 9.07 & 9.63 & $-21.7$ & $-22.0$ & 0.3 \\
May 20, 2005   & S\_SPEC\_CO\_CIRS\_1495279701\_FP1 & 9.07 & 9.63 & $-21.7$ & $-21.8$ & 9.3 \\
\hline                        
January 28, 2008 & S\_SPEC\_CO\_CIRS\_1580221333\_FP1 & 9.27 & 8.40 & $-8.6$ & $-16.9$ & 27.4 \\
\hline                                   
\end{tabular}
\end{table*}

The temperature profile measured by COMICS in 2008 is compared with that by CIRS in 2008 at $\alpha = 27.4^\circ$. 
Although global trend of temperature profile by COMICS and CIRS is similar to each other, COMICS measurements provide significantly higher temperatures than CIRS. 
At a lower solar elevation above the ring plane, Saturn's ring temperatures show more notable dependence on the phase angle. 
The dependence is interpreted as shadowing effect of particles in the rings---larger shadowed area in the rings appears to the observer at a larger phase angle, while no shadows would be seen at $\alpha \sim 0$. 
At a lower solar elevation, shadowed fractional area in the rings is larger \citep[e.g.][]{froidevaux81, flandes10}, and thus the dependence of the ring temperatures on the phase angle is stronger than at a higher elevation. 
Our COMICS images of Saturn's rings presented in this work are complementary to CIRS observations, which did not cover very low phases ($\alpha < 20^\circ$) at low solar elevations ($|B'| < 10^\circ$). 
Future application of surface model of Saturn's rings and ring particles \citep{morishima16} to our COMICS data would give us an insight into the influences of intra-particle shadows caused by surface roughness of individual particles and inter-particle shadowing to the physical properties in the rings at various geometries.

\section{Summary}

Multi-band MIR images of Saturn's rings in 2008 and 2005 observed by Subaru/COMICS were presented. We found that the C ring and the Cassini Division were warmer than the B and A rings in 2008. The C ring and the Cassini Division have lower albedos, lower optical depths, and smaller self-shadowing effect than the B and A rings, all of which are likely to contribute to the higher temperatures in the C ring and the Cassini Division. We also found that the radial profile of the MIR emission contrast of Saturn's rings in 2008 was the inverse of that in 2005, and that the C ring and the Cassini Division were much brighter than the B and A rings in 2008. This temporal variation is probably caused by seasonal changes of the elevations of the sun and observer above the ring plane as varying angles will lead to differing filling factors and temperatures of the particles in the rings.

\begin{acknowledgements}
This research is based on data collected at Subaru Telescope and obtained from the SMOKA, which is operated by the Astronomy Data Center, National Astronomical Observatory of Japan. We appreciate the referee, L.\,Spilker, for her comments which have greatly helped to improve the manuscript. We are deeply grateful to P.A.\,Yanamandra-Fisher, G.S.\,Orton, and L.N.\,Fletcher, the core members of the original observing team with COMICS, who gave us very helpful comments, and also to the other members, B.\,Fisher, T.\,Fuse, T.\,Greathouse, N.\,Takato, H.\,Kawakita, R.\,Furusho, and S.\,Takeuchi. 
We thank H.\,Kobayashi, D.\,Kinoshita, and T.\,Onaka for their suggestions and supports. This work benefited from fruitful discussions with A.T.\,Tokunaga, T.\,Kostiuk, and other participants at the ESO workshop ``Ground and space observatories: a joint venture to planetary science'' in March 2015, chaired by E.\,Villard and O.\,Witasse.
This work was supported by JSPS KAKENHI Grant Numbers JP23103002 and JP26800110. 
\end{acknowledgements}



\bibliographystyle{aa} 
\bibliography{references.bib} 

\newcommand{\noop}[1]{}
\begin{thebibliography}{37}
\expandafter\ifx\csname natexlab\endcsname\relax\def\natexlab#1{#1}\fi

\bibitem[{{Allen} \& {Murdock}(1971)}]{allen71}
{Allen}, D.~A. \& {Murdock}, T.~L. 1971, \icarus, 14, 1

\bibitem[{{Altobelli} {et~al.}(2007){Altobelli}, {Spilker}, {Pilorz}, {Brooks},
  {Edgington}, {Wallis}, \& {Flasar}}]{altobelli07}
{Altobelli}, N., {Spilker}, L., {Pilorz}, S., {et~al.} 2007, \icarus, 191, 691

\bibitem[{{Altobelli} {et~al.}(2008){Altobelli}, {Spilker}, {Leyrat}, \&
  {Pilorz}}]{altobelli08}
{Altobelli}, N., {Spilker}, L.~J., {Leyrat}, C., \& {Pilorz}, S. 2008, \planss,
  56, 134

\bibitem[{{Canup}(2010)}]{canup10}
{Canup}, R.~M. 2010, \nat, 468, 943

\bibitem[{{Charnoz} {et~al.}(2009){Charnoz}, {Morbidelli}, {Dones}, \&
  {Salmon}}]{charnoz09}
{Charnoz}, S., {Morbidelli}, A., {Dones}, L., \& {Salmon}, J. 2009, \icarus,
  199, 413

\bibitem[{{Cohen} {et~al.}(1999){Cohen}, {Walker}, {Carter}, {Hammersley},
  {Kidger}, \& {Noguchi}}]{cohen99}
{Cohen}, M., {Walker}, R.~G., {Carter}, B., {et~al.} 1999, \aj, 117, 1864

\bibitem[{{Cohen} {et~al.}(1995){Cohen}, {Witteborn}, {Walker}, {Bregman}, \&
  {Wooden}}]{cohen95}
{Cohen}, M., {Witteborn}, F.~C., {Walker}, R.~G., {Bregman}, J.~D., \&
  {Wooden}, D.~H. 1995, \aj, 110, 275

\bibitem[{{Colwell} {et~al.}(2010){Colwell}, {Esposito}, {Jerousek}, {Srem{\v
  c}evi{\'c}}, {Pettis}, \& {Bradley}}]{colwell10}
{Colwell}, J.~E., {Esposito}, L.~W., {Jerousek}, R.~G., {et~al.} 2010, \aj,
  140, 1569

\bibitem[{{Colwell} {et~al.}(2009){Colwell}, {Nicholson}, {Tiscareno},
  {Murray}, {French}, \& {Marouf}}]{colwell09}
{Colwell}, J.~E., {Nicholson}, P.~D., {Tiscareno}, M.~S., {et~al.} 2009, {The
  Structure of Saturn's Rings}, ed. M.~K. {Dougherty}, L.~W. {Esposito}, \&
  S.~M. {Krimigis}, 375

\bibitem[{{Cuzzi} \& {Estrada}(1998)}]{cuzzi98}
{Cuzzi}, J.~N. \& {Estrada}, P.~R. 1998, \icarus, 132, 1

\bibitem[{{Cuzzi} {et~al.}(2002){Cuzzi}, {French}, \& {Dones}}]{cuzzi02}
{Cuzzi}, J.~N., {French}, R.~G., \& {Dones}, L. 2002, \icarus, 158, 199

\bibitem[{{Esposito}(2011)}]{esposito11}
{Esposito}, L.~W. 2011, {Planetary Rings}

\bibitem[{{Ferrari} {et~al.}(2005){Ferrari}, {Galdemard}, {Lagage}, {Pantin},
  \& {Quoirin}}]{ferrari05}
{Ferrari}, C., {Galdemard}, P., {Lagage}, P.~O., {Pantin}, E., \& {Quoirin}, C.
  2005, \aap, 441, 379

\bibitem[{{Flandes} {et~al.}(2010){Flandes}, {Spilker}, {Morishima}, {Pilorz},
  {Leyrat}, {Altobelli}, {Brooks}, \& {Edgington}}]{flandes10}
{Flandes}, A., {Spilker}, L., {Morishima}, R., {et~al.} 2010, \\\planss, 58,
  1758

\bibitem[{{Fletcher} {et~al.}(2009){Fletcher}, {Orton}, {Yanamandra-Fisher},
  {Fisher}, {Parrish}, \& {Irwin}}]{fletcher09}
{Fletcher}, L.~N., {Orton}, G.~S., {Yanamandra-Fisher}, P., {et~al.} 2009,
  \icarus, 200, 154

\bibitem[{{French} \& {Nicholson}(2000)}]{french00}
{French}, R.~G. \& {Nicholson}, P.~D. 2000, \icarus, 145, 502

\bibitem[{{Froidevaux}(1981)}]{froidevaux81}
{Froidevaux}, L. 1981, \icarus, 46, 4

\bibitem[{{Froidevaux} {et~al.}(1981){Froidevaux}, {Matthews}, \&
  {Neugebauer}}]{froidevaux81a}
{Froidevaux}, L., {Matthews}, K., \& {Neugebauer}, G. 1981, \icarus, 46, 18

\bibitem[{{Fujiwara} {et~al.}(2013){Fujiwara}, {Ishihara}, {Onaka}, {Takita},
  {Kataza}, {Yamashita}, {Fukagawa}, {Ootsubo}, {Hirao}, {Enya}, {Marshall},
  {White}, {Nakagawa}, \& {Murakami}}]{fujiwara13}
{Fujiwara}, H., {Ishihara}, D., {Onaka}, T., {et~al.} 2013, \aap, 550, A45

\bibitem[{{Fujiwara} {et~al.}(2012){Fujiwara}, {Onaka}, {Takita}, {Yamashita},
  {Fukagawa}, {Ishihara}, {Kataza}, \& {Murakami}}]{fujiwara12b}
{Fujiwara}, H., {Onaka}, T., {Takita}, S., {et~al.} 2012, \apjl, 759, L18

\bibitem[{{Kataza} {et~al.}(2000){Kataza}, {Okamoto}, {Takubo}, {Onaka},
  {Sako}, {Nakamura}, {Miyata}, \& {Yamashita}}]{kataza00}
{Kataza}, H., {Okamoto}, Y., {Takubo}, S., {et~al.} 2000, in Presented at the
  Society of Photo-Optical Instrumentation Engineers (SPIE) Conference, Vol.
  4008, Society of Photo-Optical Instrumentation Engineers (SPIE) Conference
  Series, ed. {M.~Iye \& A.~F.~Moorwood}, 1144--1152

\bibitem[{{Leyrat} {et~al.}(2008{\natexlab{a}}){Leyrat}, {Ferrari}, {Charnoz},
  {Decriem}, {Spilker}, \& {Pilorz}}]{leyrat08b}
{Leyrat}, C., {Ferrari}, C., {Charnoz}, S., {et~al.} 2008{\natexlab{a}},
  \icarus, 196, 625

\bibitem[{{Leyrat} {et~al.}(2008{\natexlab{b}}){Leyrat}, {Spilker},
  {Altobelli}, {Pilorz}, \& {Ferrari}}]{leyrat08a}
{Leyrat}, C., {Spilker}, L.~J., {Altobelli}, N., {Pilorz}, S., \& {Ferrari}, C.
  2008{\natexlab{b}}, \planss, 56, 117

\bibitem[{{Lord}(1992)}]{lord92}
{Lord}, S.~D. 1992, {A new software tool for computing Earth's atmospheric
  transmission of near- and far-infrared radiation}, Tech. rep.

\bibitem[{{Lynch} {et~al.}(2000){Lynch}, {Mazuk}, {Russell}, {Hackwell}, \&
  {Hanner}}]{lynch00}
{Lynch}, D.~K., {Mazuk}, A.~L., {Russell}, R.~W., {Hackwell}, J.~A., \&
  {Hanner}, M.~S. 2000, \icarus, 146, 43

\bibitem[{{Morishima} {et~al.}(2009){Morishima}, {Salo}, \&
  {Ohtsuki}}]{morishima09}
{Morishima}, R., {Salo}, H., \& {Ohtsuki}, K. 2009, \icarus, 201, 634

\bibitem[{{Morishima} {et~al.}(2011){Morishima}, {Spilker}, \&
  {Ohtsuki}}]{morishima11}
{Morishima}, R., {Spilker}, L., \& {Ohtsuki}, K. 2011, \icarus, 215, 107

\bibitem[{{Morishima} {et~al.}(2010){Morishima}, {Spilker}, {Salo}, {Ohtsuki},
  {Altobelli}, \& {Pilortz}}]{morishima10}
{Morishima}, R., {Spilker}, L., {Salo}, H., {et~al.} 2010, \icarus, 210, 330

\bibitem[{{Morishima} {et~al.}(2016){Morishima}, {Turner}, \&
  {Spilker}}]{morishima16}
{Morishima}, R., {Turner}, N., \& {Spilker}, L. 2016, \icarus, submitted

\bibitem[{{Nolt} {et~al.}(1978){Nolt}, {Tokunaga}, {Gillett}, \&
  {Caldwell}}]{nolt78}
{Nolt}, I.~G., {Tokunaga}, A., {Gillett}, F.~C., \& {Caldwell}, J. 1978, \apjl,
  219, L63

\bibitem[{{Okamoto} {et~al.}(2003){Okamoto}, {Kataza}, {Yamashita}, {Miyata},
  {Sako}, {Takubo}, {Honda}, \& {Onaka}}]{okamoto02}
{Okamoto}, Y.~K., {Kataza}, H., {Yamashita}, T., {et~al.} 2003, in Presented at
  the Society of Photo-Optical Instrumentation Engineers (SPIE) Conference,
  Vol. 4841, Society of Photo-Optical Instrumentation Engineers (SPIE)
  Conference Series, ed. {M.~Iye \& A.~F.~M.~Moorwood}, 169--180

\bibitem[{{Rieke}(1975)}]{rieke75}
{Rieke}, G.~H. 1975, \icarus, 26, 37

\bibitem[{{Sako} {et~al.}(2003){Sako}, {Okamoto}, {Kataza}, {Miyata}, {Takubo},
  {Honda}, {Fujiyoshi}, {Onaka}, \& {Yamashita}}]{sako03}
{Sako}, S., {Okamoto}, Y.~K., {Kataza}, H., {et~al.} 2003, \pasp, 115, 1407

\bibitem[{{Spilker} {et~al.}(2003){Spilker}, {Ferrari}, {Cuzzi}, {Showalter},
  {Pearl}, \& {Wallis}}]{spilker03}
{Spilker}, L., {Ferrari}, C., {Cuzzi}, J.~N., {et~al.} 2003, \planss, 51, 929

\bibitem[{{Spilker} {et~al.}(2006){Spilker}, {Pilorz}, {Wallis}, {Pearl},
  {Cuzzi}, {Brooks}, {Altobelli}, {Edgington}, {Showalter}, {Michael Flasar},
  {Ferrari}, \& {Leyrat}}]{spilker06}
{Spilker}, L.~J., {Pilorz}, S.~H., {Wallis}, B.~D., {et~al.} 2006, \planss, 54,
  1167

\bibitem[{{Tokunaga} {et~al.}(1980){Tokunaga}, {Caldwell}, \&
  {Nolt}}]{tokunaga80}
{Tokunaga}, A.~T., {Caldwell}, J., \& {Nolt}, I.~G. 1980, \nat, 287, 212

\bibitem[{{Verbiscer} {et~al.}(2009){Verbiscer}, {Skrutskie}, \&
  {Hamilton}}]{verbiscer09}
{Verbiscer}, A.~J., {Skrutskie}, M.~F., \& {Hamilton}, D.~P. 2009, \nat, 461,
  1098

\end{thebibliography}






\end{document}